\documentclass[12pt,preprint]{aastex63}

\shorttitle{Cosmic Ray Hits on IRAC Detectors}
\shortauthors{Hagan et al.}

\begin{document}

\title{Cosmic-Ray-Related Signals from Detectors in Space: the Spitzer/IRAC Si:As IBC Devices}

\author{J. Brendan Hagan}\affiliation{Space Telescope Science Institute, 3700 San Martin Drive, Baltimore, MD 21218, USA; now at Starfish Space Inc., Kent, WA}
\author{George Rieke}\affiliation{Steward Observatory, University of Arizona, Tucson, AZ 85721, USA}
\author{Ori D. Fox}\affiliation{Space Telescope Science Institute, 3700 San Martin Drive, Baltimore, MD 21218, USA}
\author{Alberto Noriega-Crespo}\affiliation{Space Telescope Science Institute, 3700 San Martin Drive, Baltimore, MD 21218, USA} 
\author{Dean C. Hines}\affiliation{Space Telescope Science Institute, 3700 San Martin Drive, Baltimore, MD 21218, USA} 
\author{Misty Cracraft}\affiliation{Space Telescope Science Institute, 3700 San Martin Drive, Baltimore, MD 21218, USA} 
\author{Macarena Garc{\'{\i}}a Mar{\'{\i}}n}\affiliation{European Space Agency, 3700 San Martin Drive, Baltimore, MD 21218, USA}

\email{ j.brendan.hagan@gmail.com; grieke@arizona.edu}

\begin{abstract}
We evaluate the hit rate of cosmic rays and their daughter particles on the Si:As IBC detectors in the IRAC instrument on the  Spitzer Space Telescope. The hit rate follows the ambient proton flux closely, but the hits occur at more than twice the rate expected just from this flux. Toward large amplitudes, the size distribution of hits by single-charge particles (muons) follows the Landau Distribution. The amplitudes of the hits are distributed to well below the energy loss of a traditional ``average minimum-ionizing proton'' as a result of statistical fluctuations in the ionization loss within the detectors. Nonetheless, hits with amplitudes less than a few hundred electrons are rare; this places nearly all hits in an amplitude range that is readily identified given the read noises of modern solid-state detectors. The spread of individual hits over multiple pixels is dominated by geometric effects, i.e., the range of incident angles, but shows a modest excess probably due to: (1) showering and scattering of particles; (2) the energy imparted on the ionization products by the energetic protons; and (3) interpixel capacitance. Although this study is focused on a specific detector type, it should have general application to operation of modern solid-state detectors in space.
 
\end{abstract}


\section{Introduction}
\label{sec:intro}

 There is an increasing emphasis on placing space astronomy missions in either an Earth-trailing, Solar orbit (e.g., Spitzer Space Telescope, hereafter Spitzer) or at a Lagrangian point, usually Sun-Earth L2 (e.g., Herschel Space Observatory (Herschel), Gaia, and the James Webb Space Telescope (JWST)). In such locations, the energetic particle  environment consists of Solar protons (and heavier nuclei) and Galactic Cosmic Rays (GCRs), which are also mainly (89\%) protons, but can include helium (10\%) and heavier nuclei (1\%). With modest shielding to eliminate Solar protons, the environment of energetic particles is then dominated by GCRs, which produce clouds of ionized particles in any detectors that they hit plus secondary particles from interactions with the material surrounding the detectors, e.g. in the Focal Plane Assemblies (FPAs)(Pickel et al. 2002). Predicting and modeling the performance of such space missions depends on a thorough understanding of these spurious signals for the specific detector type employed. Information about this behavior has been summarized for the bolometers on Herschel \citep{horeau2012}, the CCDs on Gaia \citep{crowley2016}, the detectors on the Infrared Space Observatory (ISO) \citep{heras1998, heras2003}, the Si:As IBC devices on Akari \citep{mouri2011}, and for NICMOS on the Hubble Space Telescope (HST) \citep{ladbury2002}. Preliminary studies are also available for the Spitzer IRAC detectors \citep{patten2004,hora2006}.

This investigation extends the studies of GCR effects on the {\it Spitzer} Infrared Array Camera (IRAC) Si:As IBC detector arrays in Channels 3 and 4 (5.8 and 8.0 \micron, respectively). Although we focus on this specific type of detector, our investigation illustrates many general characteristics of the reaction of solid-state detectors to GCRs. Spitzer was placed into an Earth-trailing, Solar orbit, which positions the spacecraft well outside any shielding effects of the Earth's magnetosphere or any enhancements in charged particle flux due to the Earth's radiation belts. The resulting environment should be very similar to that for spacecraft at the Sun-Earth L2 Lagrangian Point, such as Herschel, Gaia, and JWST. 

Specific issues to be addressed in this paper include: (1) the distribution of numbers of ionized particles produced in GCR hits on the detectors, and specifically the origin of relatively small signals; (2) the excess in the cosmic ray hit rate on many missions compared with predictions from models -- we show that this is also the case for the IRAC detectors; (3) effects that disturb detector pixels beyond the minimum that are hit directly as a result of the angle of incidence of an energetic particle -- effects such as interpixel capacitance, scattering and showering within the detector array, and propagation of particles given energy kicks by the incident particle.

Section \ref{sec:data} describes the IRAC Si:As IBC data that we analyze as described in Section \ref{sec:methods}. The results are described in Section \ref{sec:results} in terms of the overall behavior, i.e., the relation of the hit rate to the proton flux, and the intensity amplitude spectrum of the hits. This discussion raises some hypotheses about the nature of the individual hits, which we discuss in Section~\ref{sec:hitchar}. Our results are summarized along with recommendations for cosmic ray hit mitigation in Section \ref{sec:conclusion}.

\section{Data Description}
\label{sec:data}

\subsection{{\it Spitzer} IRAC Data}

 The Si:As IBC detector arrays in IRAC have a format of $256 \times 256$ pixels with each pixel of dimensions $30 \times 30$ $\mu$m, under which is an infrared-active layer $25$ $\mu$m thick followed by a $4$ $\mu$m thick blocking layer followed by the output electrical contact. This structure is back-illuminated through a $\sim$ 500 $\mu$m thick substrate with a buried electrical contact between the substrate and IR-active layer. 

\begin{figure}
\centering
 \fbox{\includegraphics[width=0.45\textwidth]{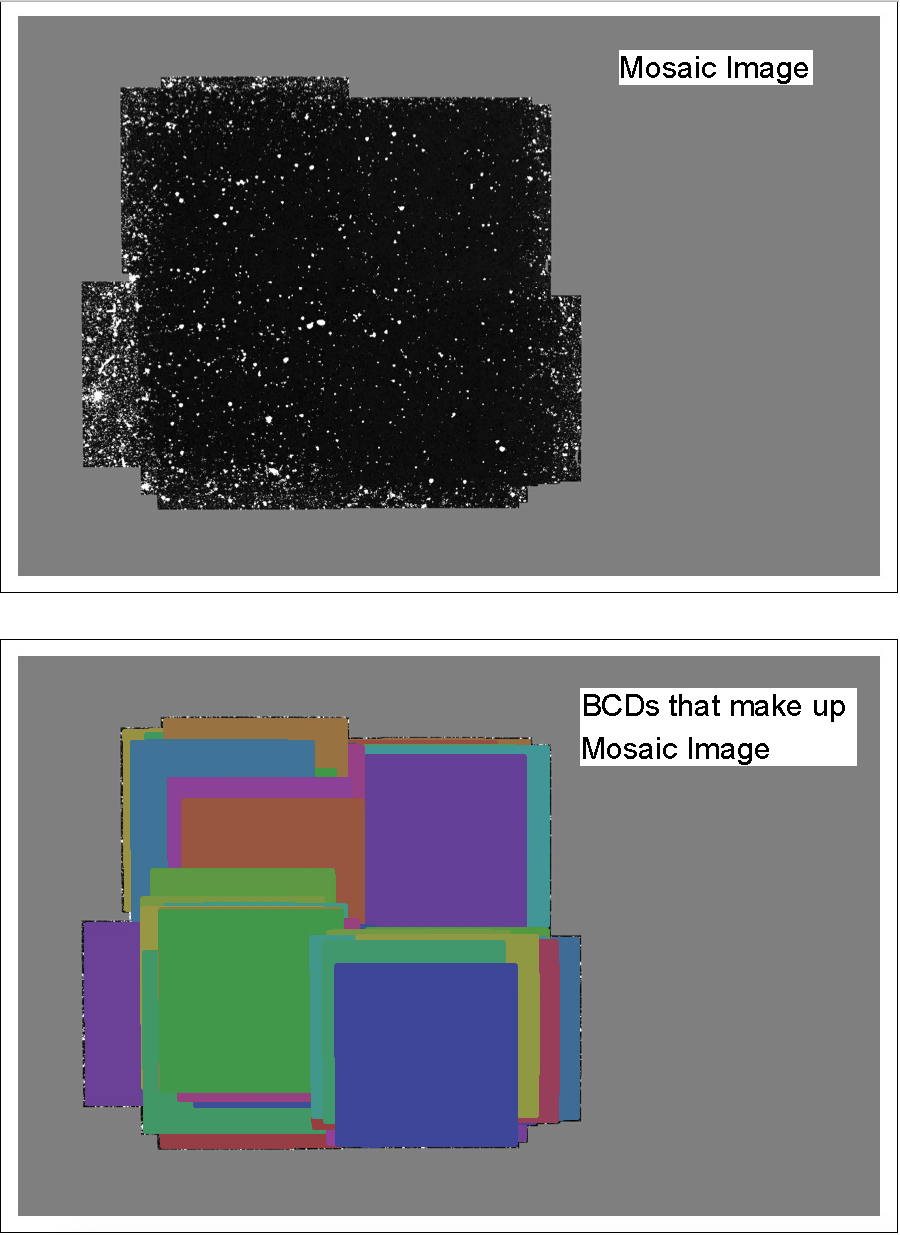}}
 \caption{A mosaic file (top), and a schematic representation of the 50 BCD files used to make the mosaic file (bottom).}
 \label{fig1}
\end{figure}

We retrieved both Level 1 and Level 2 data products obtained with the IRAC arrays from the Spitzer Heritage Archive\footnote{{\url{http://sha.ipac.caltech.edu}}}. The Level 1 products contain Basic Calibrated Data (BCD) files, which are individual exposures that have not been corrected for outlier rejection. The Level 2 products contain the mosaic files (*maic.fits), which were resampled, and weighted images derived from the BCDs excluding outlier-affected pixels. Figure \ref{fig1} shows an example of a mosaic image (top), and the coverage of the BCD files (bottom) that were used to make the mosaic image.

\begin{figure}
\centering
\includegraphics[width=0.85\textwidth]{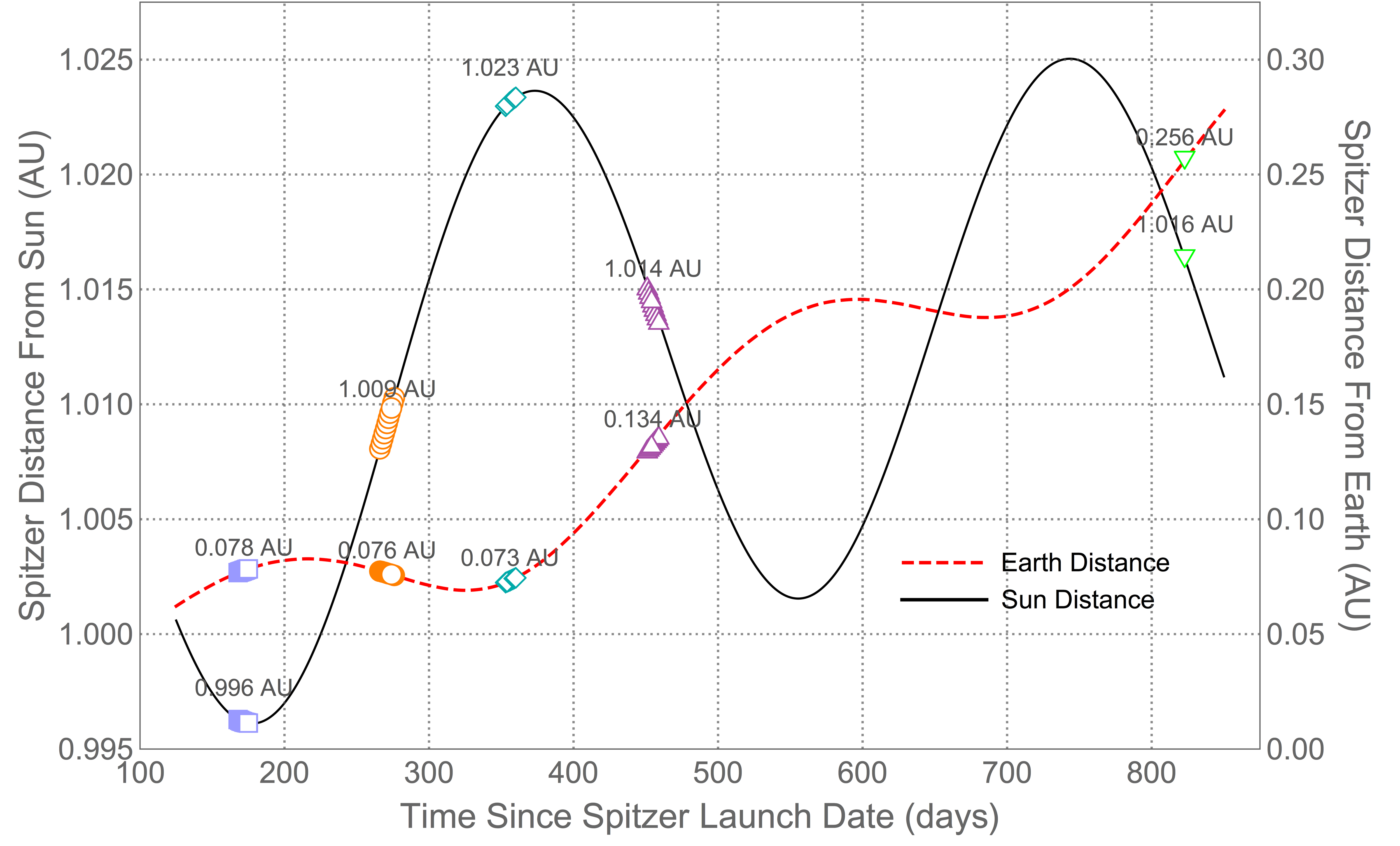}
\caption{Ephemeris data for the {\it Spitzer Space Telescope} with respect to the Earth and Sun. The colored markers show time during which the  observations we analyze were taken.\label{fig2}}
\end{figure}

\begin{figure}
\centering
\includegraphics[width=0.8\textwidth]{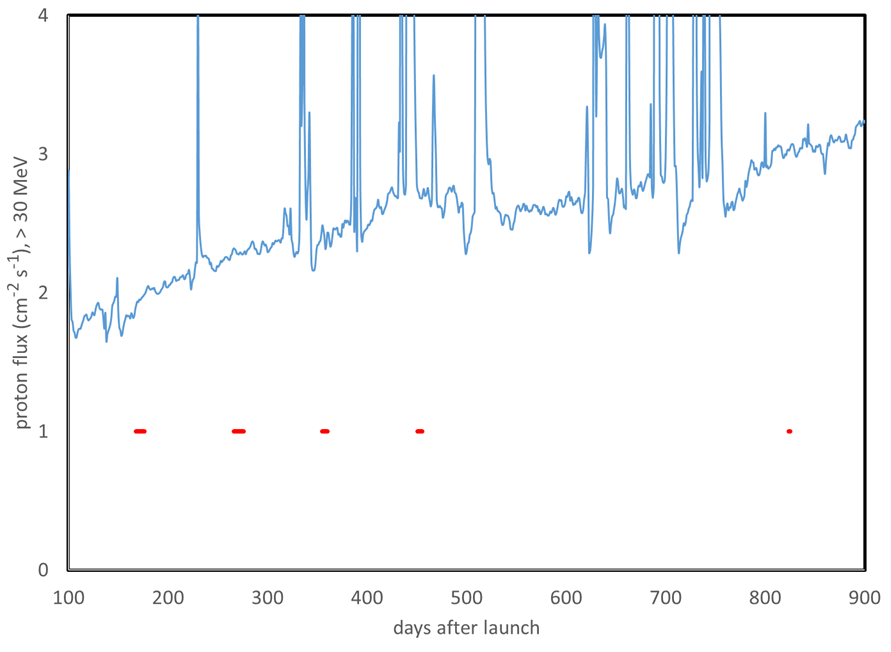}
\caption{Observed flux of protons at $>$ 30 MeV (continuous blue line) derived in Appendix A and times used for our analysis (short red bars). \label{fig3}}
\end{figure}

Table \ref{tab1} summarizes the data used for this study.  PIDs 169 and 194 (GOODS North and South) are ideal because they provide long, uninterrupted exposures in low background regions.  Figure \ref{fig2} plots the spacecraft position in distance from the Sun and Earth as a function of observation date. Even by the first epoch, the spacecraft is well beyond the distance of the L2 Lagrange point ($\sim$ 0.01 AU from Earth), and beyond any shielding effects from Earth's magnetosphere. The primary shielding is thus provided by the spacecraft and instrument structure. An aluminum thickness of 4 mm is sufficient to block nearly all of the proton flux (i.e., protons $<$ 20 MeV) from the quiescent sun. However, coronal mass ejections accelerate protons into the hundreds of MeV range \citep[e.g.,][]{bruno2019}, and during these events the Solar proton flux can penetrate the shielding to dominate the flux received at the detectors. Figure \ref{fig3} shows the time periods of our analysis compared with the measured overall proton flux (the details of the flux derivation are provided in  Appendix A). Our analysis focuses on quiet periods, and hence the proton events are dominated by GCRs, the flux of which varies in anti-synchronism with the solar cycle (i.e., is lowest near solar max).  .

\begin{table}
\scriptsize
\centering
\caption{Observation Details}
\begin{tabular}{llcccc}
\hline
Epoch & PIDs & PI & Total Exposure Time & Spitzer Distance from Earth & Solar Activity\\
(days since launch) & & & Ch3/Ch4 (sec) & (AU) & (Solar Flux Units)\\
\hline
168-176 & 169, 194 & Dickinson & 325,000/325,000 & 0.078 & 110 (10) \\
266-276 & 169, 194 & Dickinson & 505,000/505,000 & 0.076 & 105 (10) \\
355-360 & 169, 194 & Dickinson & 300,000/300,000 & 0.073 & 138 (10) \\
451-455 & 169, 194 & Dickinson & 480,000/480,000 & 0.134 & 105 (5) \\
824-825 & 169, 194 & Dickinson &  10,000/10,000 & 0.256 & 85 (5) \\
\end{tabular}

\label{tab1}
\end{table}

\subsection{MIRI Detector Data}

 Comparison with data from ground test of the Mid-Infrared Instrument (MIRI) for JWST can provide a useful perspective. The baseline MIRI detector arrays are similar to the IRAC arrays except they are $1024 \times 1024$ pixels in format and each pixel is $25 \times 25$ $\mu$m, with an IR-active layer $35 \mu$m thick. On the ground, cosmic ray hits on a detector array are dominated by energetic muons, because these particles penetrate the atmosphere. Thus, these events show the effects of hits by particles exclusively with a single atomic charge. This can be compared directly with results for the IRAC arrays in space, because the MIRI Si:As IBC detectors are very similar to those in IRAC (they have a much larger format, but otherwise with 20\% smaller pixels but a 40\% thicker IR-active layer, the pixel areas (which are of primary interest for this study) are virtually identical) \citep{ressler2008, rieke2015}. Data for these arrays were obtained from the third instrument cryogenic-vacuum instrument testing (CV3). 

\section{Methods for Data Analysis}
\label{sec:methods}

\subsection{IRAC Data}

To identify and measure CR hits, we perform the same analysis on all of the IRAC data.  Each BCD file has a corresponding Rmask file (*brmsk.fits) that flags outlier pixels. For our datasets, the Rmasks were all generated with box outlier rejection where an outlier pixel was marked with a bit 3 value, or value = 8, as described in the MOPEX User Guide \citep{makovoz2012}. The examined data sets had robust outlier rejection for each BCD, since there was sufficient coverage to support this process. To obtain reliable results for this work, we inspect the individual Rmask files to exclude poor and/or unreliable masks.  For example,
we ignore noisy columns and rows along the array edges (following \citet{hora2006}).
Specifically, we examine a 246x246 area on the array, ignoring the first and last 5 rows and columns of the detector. 

Next, we resample the BCD and mosaic files so that they have the same pixel size to allow for direct flux comparisons.  To remove the background in each field, the median value of each BCD and mosaic image is subtracted from the image itself (we will refer to this as the ``leveled BCD image'' and ``leveled mosaic region,'' respectively).  The pixel values of CR hits identified in the Rmasks are first measured in the individual ``leveled BCD images.''  To remove any real flux from underlying astronomical features, we then subtract the corresponding values (same point in the sky) in the ``leveled mosaic region'', which have had any potential CR contributions removed in the mosaic and stacking process. The resulting values are the measured intensities of the radiation hits used throughout this paper. Ultimately, only 16 Rmasks were rejected from the Channel 3 data set, and 3 Rmasks were rejected from the Channel 4 data set. The total number of BCDs analyzed is 8509 and 34094 for Channels 3 and 4, respectively.

\subsection{MIRI Data}

The MIRI data, full-frame readouts in FAST mode, were reduced through a standard pipeline to produce smooth, unstructured images. 
We then identified ``jumps'' at a threshold of ten standard deviations above the noise level for a pixel.\footnote{\url{https://jwst-pipeline.readthedocs.io/en/latest/jwst/jump/description.html##algorithm}}
To eliminate instances where these jumps might not be cosmic ray hits, we rejected pixels with elevated noise levels, including: (1) cases identified in the first three reads of an integration ramp, when the MIRI arrays are recovering from the reset; 
(2) instances occurring within 10 pixels of the edge of the array; 
and (3) any jumps in the vicinity of regions of bad pixels (high dark current, excess noise etc.).
The amplitude of the accepted jumps was determined by fitting straight lines to the signal prior to and after the cosmic ray event. The size of the event was measured by counting the affected pixels (with jumps similarly identified) within a $51 \times 51$~pixel box.     

\section{Overall Radiation Hit Results}
\label{sec:results}

Table \ref{tableNumbers} summarizes the overall rates in different units. For this study, a radiation hit, or ``RadHit,'' is defined to be an isolated group of flagged outlier pixels, where ``isolation'' means there are no other flagged pixels directly to the left, right, top, or bottom of the group of flagged outliers. We can therefore also define the ``size'' of the radiation hit to be the number of flagged pixels that are connected in this way.  We also provide the hit rate per unit area in terms of hits/sec/$\mu$m$^2$ (from which we determine the fraction of the detector being hit per second). We also consider the potential effects of ``cross-talk'', where charge in one pixel impacts the charge measured in a neighboring pixel, which we discuss in more detail in Section \ref{crosstalk_sec}.
For comparison, we also include results from other investigations of the cosmic ray hit rates for the same IRAC channels.

\subsection{Trends in hit rate versus time}

Table \ref{tableNumbers} lists, and Figure \ref{fig12} shows, the individual pixel hit rates versus time. There is a slow increase in the hit rate over the duration of the IRAC observations. This behavior is expected, because the level of Solar activity was changing \citep[e.g.,][]{ross2019}, and reflects the resulting increase in proton flux as shown in Figure~\ref{fig3}. Additionally, based on Figures \ref{fig2} and \ref{fig12}, we do not find a correlation between the hit rate and location of the {\it Spitzer} spacecraft relative to the Earth and Sun.

The scatter in the hit rate is large, significantly greater than would result if the hits were from independent events.
Furthermore, the rates are substantially larger in Channel 3 than Channel 4 (5.23 pixels/second versus 4.32 pixels per second, respectively). The difference presumably reflects differences in the degree of shielding and showering\footnote{Created from knock-on
electrons, or ``delta rays,'' resulting from transfer of energy from a high-energy  cosmic ray to electrons through interactions either with the detector material or that surrounding the detector.} for the two detector arrays.  

\begin{table*}
\centering
\caption{Cosmic Ray Measurements}
\centering
\scriptsize
\begin{tabular}{lccccl}
\hline
Detector / Analysis & Hit Rate Units & Channel 3 & Channel 4 & Area (pixels)\\
\hline
IRAC / This Work (no crosstalk)& pixel/sec  & 5.23 $\pm 0.82$  & 4.32 $\pm 1.14$ & 246x246\\
IRAC / This Work (w/ crosstalk)& pixel/sec  & 16.27 $\pm 2.13$ & 15.98  $\pm 3.65$ & 246x246\\
IRAC / This Work (no crosstalk)& hit/sec/$\mu$m$^2$  & 4.76$\times$10$^{-8}$ $\pm$ 7.47$\times$10$^{-9}$  & 3.94$\times$10$^{-8}$ $\pm$ 1.04$\times$10$^{-8}$ & N/A\\
IRAC / This Work (w/ crosstalk)& hit/sec/$\mu$m$^2$  & 1.48$\times$10$^{-7}$ $\pm$ 1.94$\times$10$^{-8}$  & 1.46 $\times$10$^{-7}$ $\pm$ 3.32$\times$10$^{-8}$ & N/A\\
IRAC / This Work & RadHit/sec  & 1.49 $\pm 0.15$ & 1.69  $\pm 0.38$ & 246x246\\ 
IRAC / \citet{patten2004} & pixel/sec  & 7  & 7 & 250x250\\
IRAC /  \citet{hora2006} & pixel/sec  & 6.2  & 5.8 & 250x250\\
\hline
\end{tabular}

\label{tableNumbers}
\end{table*}

\subsection{Comparison of Hit Rates with Proton Flux levels}

\begin{figure}
\centering
 \includegraphics[width=\textwidth]{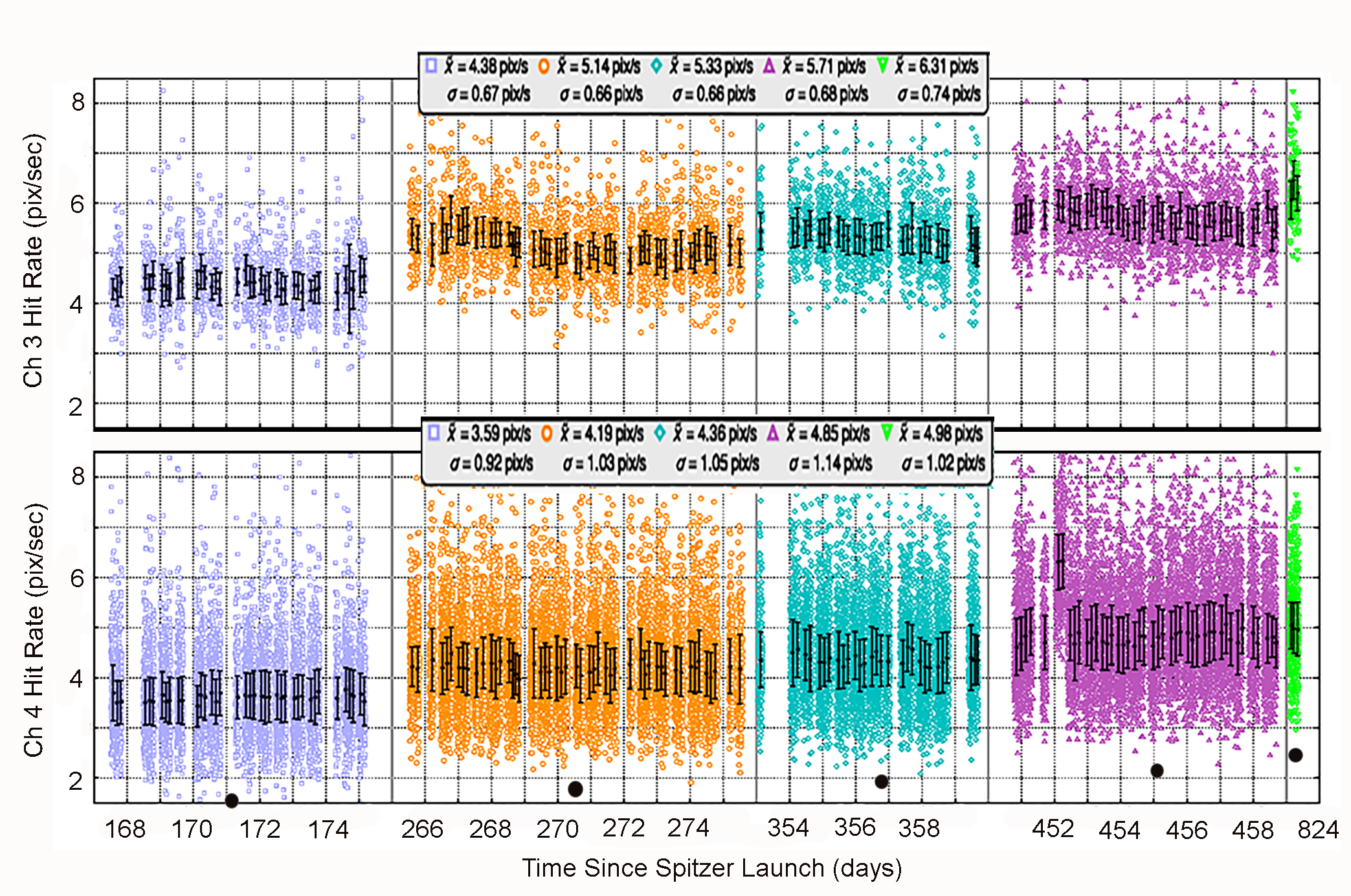}
 \caption{Hit rate vs. time. Each data point is the hit rate for a single exposure (i.e. for a single BCD file). For each collection of exposures that make up a mosaic, we plot an additional data point in black, which shows the average hit rate for that mosaic. The spread of the hit rates found in a mosaic is shown in the error bar. The x-axis is five times partitioned to show the five distinct periods of time in which the data was collected. The black dots at the bottom of the lower panel show the expected hit rates just from the proton flux in Figure \ref{fig3} - identical levels would also be expected for channel 3 (upper panel).\label{fig12}}
\end{figure}

The observed hit rates can be compared with the predicted hit rates from proton fluxes, particularly since the shielding of the detectors reduces the contribution of solar protons to far below that of the GCRs. Since the GCR energy spectrum extends to very high energies, the dependence of the predicted hit rates from them is only very weakly dependent on the details of the shielding.

The proton fluxes plotted in Figure~\ref{fig3} are isotropic, i.e., integrated over 4 $\pi$ ster. To determine the expected hit rate from the GCRs, we convert to the intensity per unit solid angle, F(GCR)/4$\pi$.  Each face of a pixel will receive hits over the equivalent of $\pi$ ster.  The surface area of an IRAC pixel is $5.28 \times 10^{-5}$ cm$^2$. We focus on a 246 $\times$ 246 pixel area of the array and compute hit rates for each of the five intervals where we analyzed the data; the lack of significant variations in the rates across each of these intervals suggests that we can ignore any time variability. The predicted hit rates are shown in Figure~\ref{fig12} as black dots across the bottom panel; the same predictions would hold for Channel 3 in the upper panel because the pixels are identical in size. It is obvious that the observed rates exceed the predictions substantially, by a factor of $\sim$ 2.7 for Channel 3 and $\sim$ 2.3 for Channel 4.

\subsection{Hit Intensity Amplitude Distribution of Electron Yields}

\begin{figure}
\centering
 \includegraphics[width=.8\textwidth]{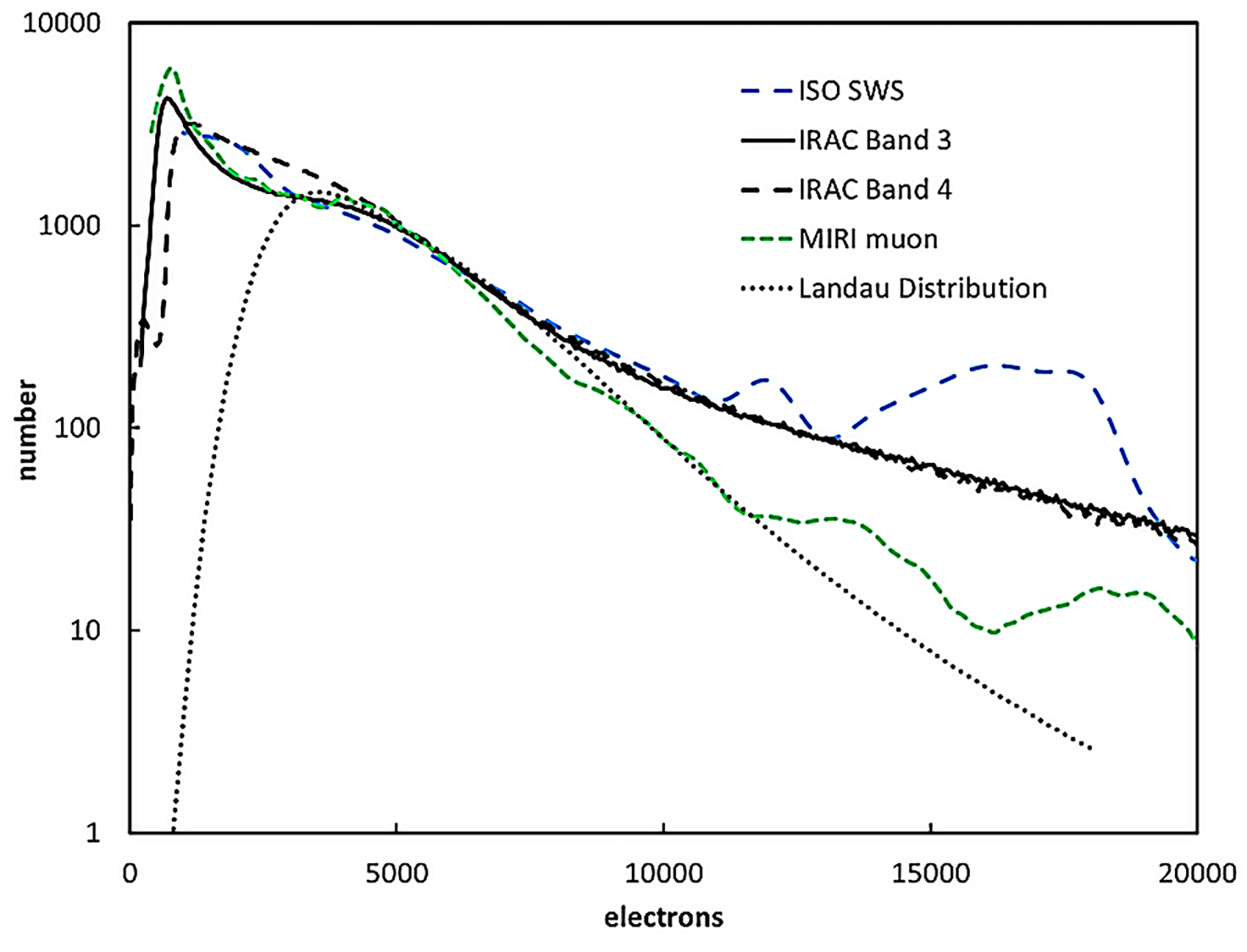}
 \caption{Single-pixel hit amplitude distribution comparison for IRAC Channels 3 and 4 and the ISO SWS detector in space and similar information for the JWST MIRI detector in the laboratory.  The Landau Distribution is the expected behavior for a single-charged particle taking account of low-probability events that cause large ionization losses. For very thin layers, it fails to account for the number of small-yield events but it is expected to account well for the rest of the size distribution \citep{physicist2021}.  The hits smaller than the peak of the distribution arise primarily from statistical fluctuations in the ionization.  The hits producing more electrons for the detectors in space are dominated by particles with more than a single atomic charge, e.g., alpha particles. } 
 \label{fig3a}
\end{figure}

\begin{figure}
\centering
 \includegraphics[width=\textwidth]{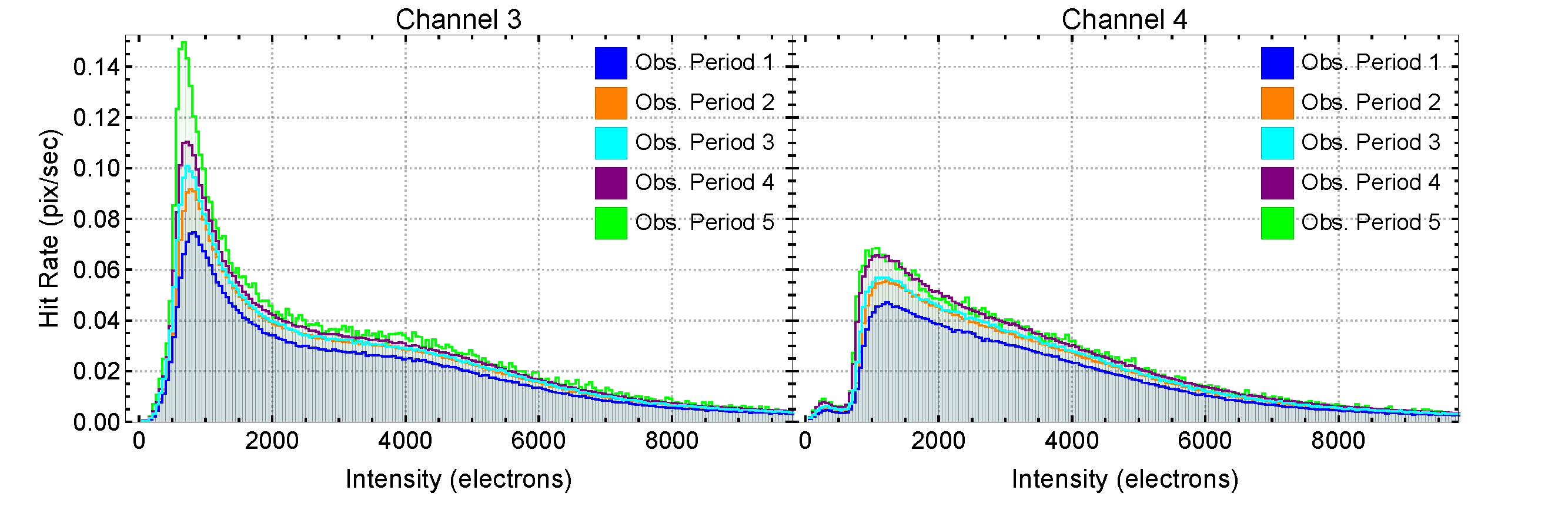}
 \caption{Hit rate as a function of the single-pixel intensity of the pixel hit, grouped by observation period, with a bin size = 50 electrons.\label{fig22}}
\end{figure}

Figure \ref{fig3a} shows the hit amplitude distribution for both the Channel 3 and 4 IRAC arrays.  Figure~\ref{fig22} plots the hit rate versus the hit amplitude.  Figure \ref{fig3a} also includes data for the  ISO Short Wavelength Spectrometer (SWS) Si:As IBC array, which were obtained from the ``SWS Glitch Tables\footnote{Obtained from \url{http://sws.ster.kuleuven.ac.be/__;!!CrWY41Z8OgsX0i-WU-0LuAcUu2o!msqEZ3a8CtO0fNMN1xKOg5zexH3iwBjni1eKFTg_9yiJ6bro6i1wfG3G1nEE2g}'' which is no longer active.},'' and for a MIRI array obtained during cryo-vacuum test of that instrument. 
The IRAC detectors have an infrared-active layer that is 25 $\mu$m thick (W. Glaccum, private communication), and a blocking layer about 4 $\mu$m thick. We could not locate the corresponding parameters for the SWS detector, but based on the time of manufacturing, the IR active layer would be about 20 $\mu$m thick (these devices were less developed than the IRAC detectors, but had similar architecture). The IR-active layer in the MIRI array is 35 $\mu$m thick. Assuming that free electrons are collected from just the IR active plus the blocking layers, the relevant detector thicknesses are $\sim$ 24, 29, and 39 \micron~for the SWS, IRAC, and MIRI detectors, respectively. 

A cosmic ray with a single atomic charge and at the minimum ionizing energy, at normal incidence onto a detector, will produce {\it on average} $\sim$ 3000 free electrons in the 29 $\mu$m thick infrared-active plus blocking layers of each of the IRAC arrays (assuming the energy loss to be the minimum, namely 1.66 Mev g$^{-1}$ cm$^{-2}$ and the average energy required to free an electron to be 3.6 eV). We describe this as the minimum average ionizing yield.  This region shows as a ``bump'' on the yield curves, particularly that for Channel 3.  The drop of the yields above this region demonstrates that the buried contact, which separates the $\sim$ 500 $\mu$m thick substrate from the IR active layer, is an impermeable barrier for free electrons produced by cosmic ray ionization in the detector substrate.  

To understand the full behavior, it is useful to start with the MIRI detector, where virtually all the hits are from muons, i.e. have single atomic charge.  In this case, the behavior should be as described by the Landau Distribution; the curve in Figure  \ref{fig3a} is the combination of two such distributions for minimum average ionizing particles, one for normal incidence and the other for incidence 45$^{\rm o}$ off normal. 
Hits with electron yields above the muon/Landau behavior towards the right of Figure \ref{fig3a} are likely to be dominated by particles with multiple atomic charges, e.g., alpha particles (ionization loss goes as atomic number squared). The three sets of data obtained in space (i.e., for the ISO and IRAC arrays), where the detectors are exposed to the full set of cosmic ray particles, all show substantially more events above the range than can be attributed to high-energy single-charge particles striking the detector array.

The excess of hits below the minimum average ionizing particle yield (towards the left of Figure \ref{fig3a}) has also been observed for the  HST NICMOS HgCdTe arrays \citep{ladbury2002}) and the Gaia CCDs \citep{crowley2016}. To first order, this behavior is due to the statistical nature of ionization loss by an energetic particle, which proceeds by discrete interactions and hence is subject to statistical fluctuations. In typical range calculations (e.g., for shielding), there are so many interactions that the statistics average out to a well determined value \citep[e.g.,][]{glasse2020}. However, the loss in the thin layers in detectors involves a small number of interactions and the range of losses is significant. This effect is frequently overlooked, but, for example, the library of simulated cosmic ray hits provided by \citet{robberto2010} includes the statistical distribution of hit sizes correctly, and the behavior is also shown in \citet{physicist2021}.

That the statistical effects are the primary source of the lack of an inflection in the hit spectrum indicative of the minimum ionizing yield is clearly demonstrated by the analysis in \citet{garcia2018} for the Gaia CCDs.
Their initial simple analytical model, which had an overly simplistic energy deposition description, was insufficient.  Instead, they carried out a Monte-Carlo simulation using the Geant4-based tool GRAS7, which is designed for radiation analysis for space. The simulation included the detectors as well as the surrounding spacecraft. The investigation also: (1) verified that protons are the dominant source of hits; and (2) and included a more complete and accurate set of energy deposition processes. The starting point for the simulation was the CREME96 cosmic ray spectrum, but the nominal shielding of the CCDs was not included since the spacecraft serves that purpose.  The more complex model confirmed that the distribution of free electron yields from cosmic ray hits on the CCDs was generally well-reproduced when the statistics of the ionization process were included.

A small number of very small hits persist above the predictions from the analysis of \citet{garcia2018} (see also \citet{mouri2011} for evidence of such behavior in Si:As IBC detectors similar to the IRAC ones).
These small hits might originate from a weakly ionizing target (i.e., the array multiplexer rather than the detectors). Alternatively, they may be unrelated to cosmic ray hits, but instead arise in momentary breakdown in high-field regions of the detector contacts \citep{rieke2002} or through other detector-related phenomena.

An important aspect of the simulated events by \citet{robberto2010} and the modeling by \citet{garcia2018} is that the apparent turnover in the numbers toward very small hits in Figure~\ref{fig3a} is real -- the hit size spectrum does not continue to grow toward smaller and smaller hits. Figure~\ref{fig22} highlights this point. Channel 3 has very few hits below $\sim$ 200 electrons; Channel 4 has a secondary small knee in this region, but it may be a detector artifact (spontaneous spiking) rather than a result of cosmic ray hits (as discussed in Section~\ref{sec:sub_hitsize}). In any case, the vast majority of cosmic ray hits are large enough to be easily identified with modern detector arrays that have read noises of 30 electrons or less.

\subsection{Behavior Under Sustained High Hit Rates}

The Spitzer Earth-trailing orbit kept the IRAC detectors away from sustained high hit rates.
However, the Akari observatory in an Earth-centered, polar orbit passed through the South Atlantic Anomaly (SAA) three or four times a day, where its virtually identical Si:As IBC detectors were exposed to rates orders of magnitude greater than the quiescent GCR rate \citep{mouri2011}. Under these conditions, the overall dark current from the entire detector array was elevated substantially \citep{mouri2011}, but it recovered to nominal values soon after the detectors were no longer in the region of intense radiation, i.e., the SAA.  

\section{Characteristics of Individual Hits}
\label{sec:hitchar}

\subsection{Sizes of the Hits}
\label{sec:sub_hitsize}

\begin{figure}
\centering
 \includegraphics[width=\textwidth]{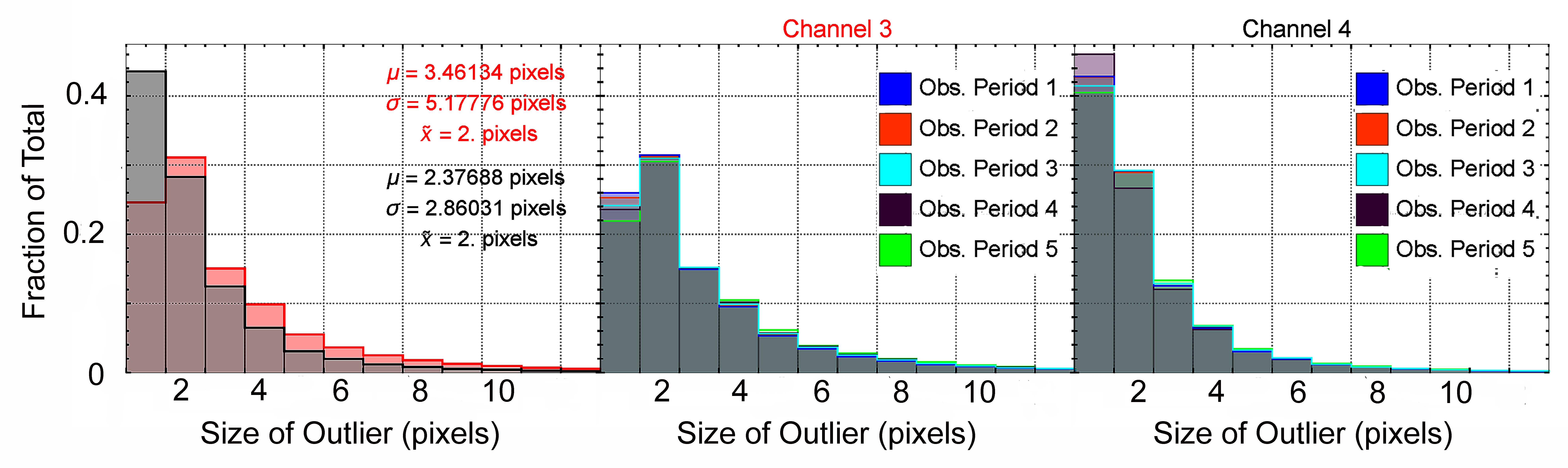}
 \caption{Size profile of potential radiation hits. Left: Comparing the profiles of Channel 3 (red) vs. Channel 4 (black-gray). Middle and Right: The pixel size profile of each observation period for data taken in Channel 3 and Channel 4 respectively.}
 \label{fig_sizes}
\end{figure}

Next we consider the size of each potential radiation hit by counting the number of flagged pixels that are in an isolated group in the Rmask image. Such groups are defined as a set of interconnected flagged pixels that do not connect to another similar set.  Figure \ref{fig_sizes} shows the size distribution of the radiation hits in number of pixels. The median size of the radiation hits in both Channels is 2 pixels. Other than the single pixel hits, the pixel size profiles appear to be similar between Channel 3 and Channel 4, though overall they are slightly larger in Channel 3. This may be because Channel 3 is operated at a lower bias voltage than Channel 4, which leads to increased charge spreading \citep{patten2004}.


\begin{figure}
\centering
 \includegraphics[width=.9\textwidth]{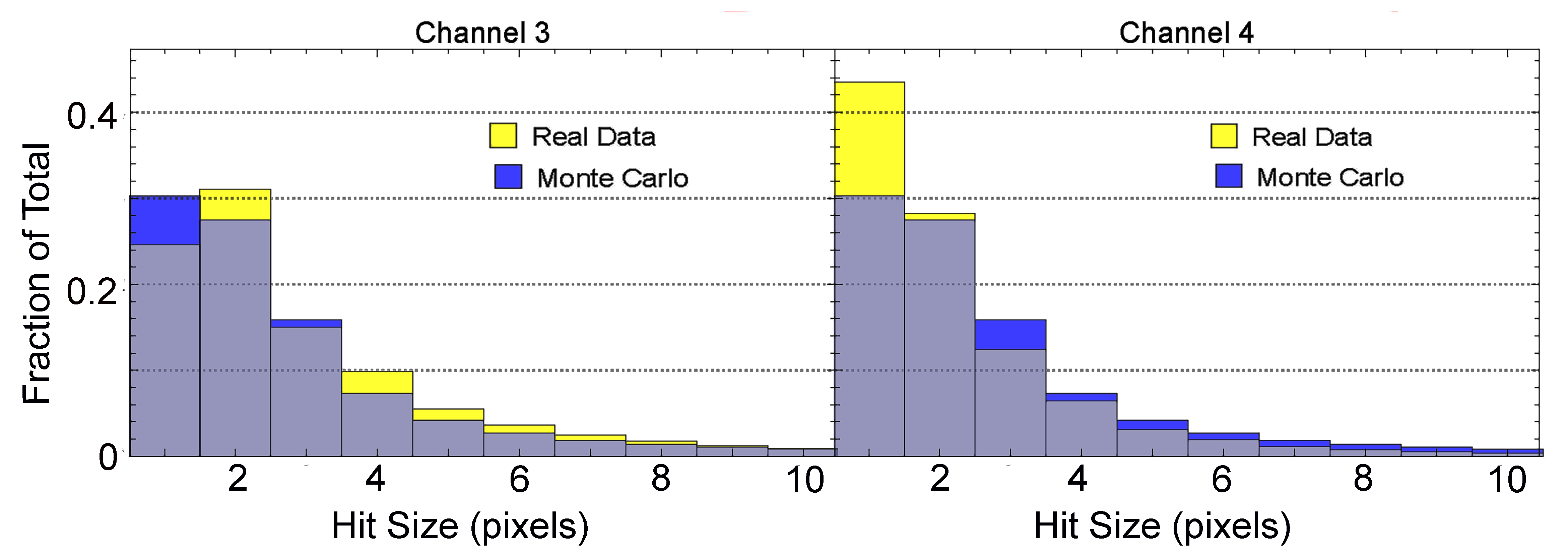}
 \caption{Monte Carlo calculation simulating an isotropic flux of  millions of cosmic rays hitting a detector with the same pixel size and height as the IRAC detector pixels (i.e. simulating cosmic rays hitting pixels with dimensions 30x30x29 \micron). }
 \label{montecarlo}
\end{figure}

Figure \ref{montecarlo} shows the results of a Monte Carlo calculation that simulates cosmic rays hitting and exiting a detector that contains pixels with same pixel physical dimensions as the IRAC pixels. This calculation gives a theoretical size profile of hits from cosmic rays incident at all angles that enter and leave an IRAC  detector. The simulation assumes no charge spreading. Each simulated cosmic ray starts at a random, normalized $(x,y)$ location $(0\leq x\leq 1, 0\leq y\leq 1)$ on the pixel surface with $z$ = $0$ and random $(\theta,\phi)$ angles $(0\leq \theta \leq \pi, 0 \leq \phi \leq 2\pi)$ to describe an angle of incidence from the isotropic flux of incoming cosmic rays.
Using the equation for a line in 3D space, we determine the location when the simulated cosmic ray leaves the detector. From here, we calculate the number of pixels the line has passed through from its start and end points. 

Figure \ref{montecarlo} compares the results from the Monte Carlo simulation with the actual data. To first order, the simulations and the data behave similarly. The most prominent discrepancy is the larger fraction of observed single-pixel hits in Channel 4 compared with predictions. This behavior is consistent with the possibility that some of these ``hits'' result from spiking in the detectors or readouts, since those events would affect single pixels only. Otherwise, the deviations of the simulation from the observed behavior appear to be in opposite directions for the two channels, i.e., an excess in the real data for Channel 3 and a deficiency for the simulation in Channel 4. However, if we assume that the single-hit count is artificially amplified in Channel 4 and redistribute the excess above the single-pixel Monte Carlo prediction into the other bins, the Channel 4 numbers exceed the Monte Carlo result just as for Channel 3. This excess above the Monte Carlo simulations probably results from: (1) scattering and showering of particles within the array or close to it \citep{patten2004}; (2) an energetic particle launching ionization products with significant energy and their ability as a result to migrate farther than the diffusion length \citep{han2017}; and (3)  crosstalk due to interpixel capacitance.

\begin{figure}
\centering
\includegraphics[width=.8\textwidth]{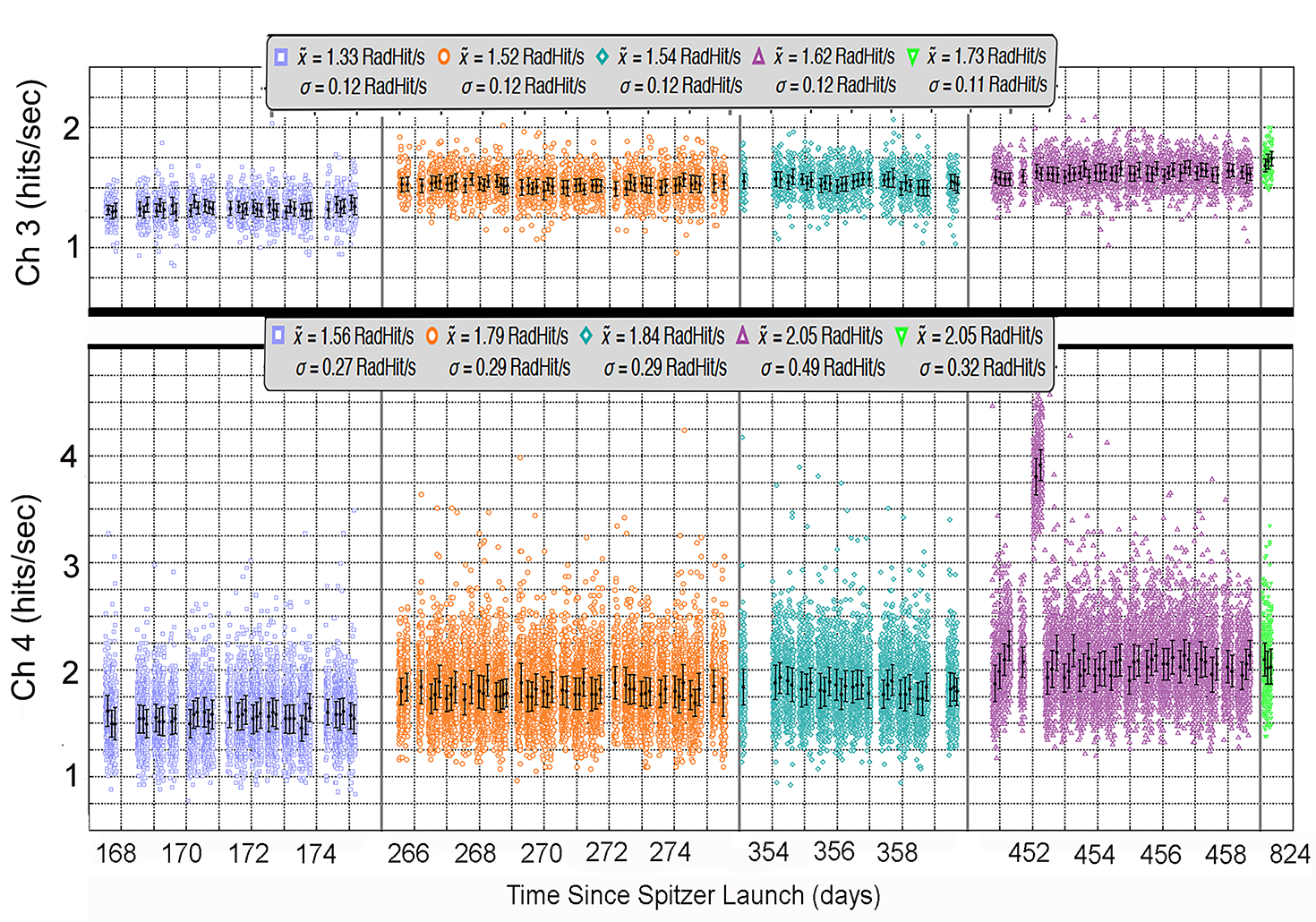}
\caption{Hit rate where one ``hit" is defined as an entire isolated group of outlier pixels.}
\label{fig26}
\end{figure}

We now re-examine the hit rate vs time for the isolated radiation hits.  While Figure \ref{fig12} plots the hit rate assuming all pixels are independent, Figure \ref{fig26} plots the hit rate assuming an isolated group is a single hit.  In this case, the hit rate is now \textit{lower} in Channel 3 than in Channel 4, which is expected because the size (in pixels) of the radiation hits in Channel 3 is larger than the radiation hits in Channel 4. We still see a trend where the hit rate increases with time. 

\begin{figure}
\centering
\includegraphics[width=.85\textwidth]{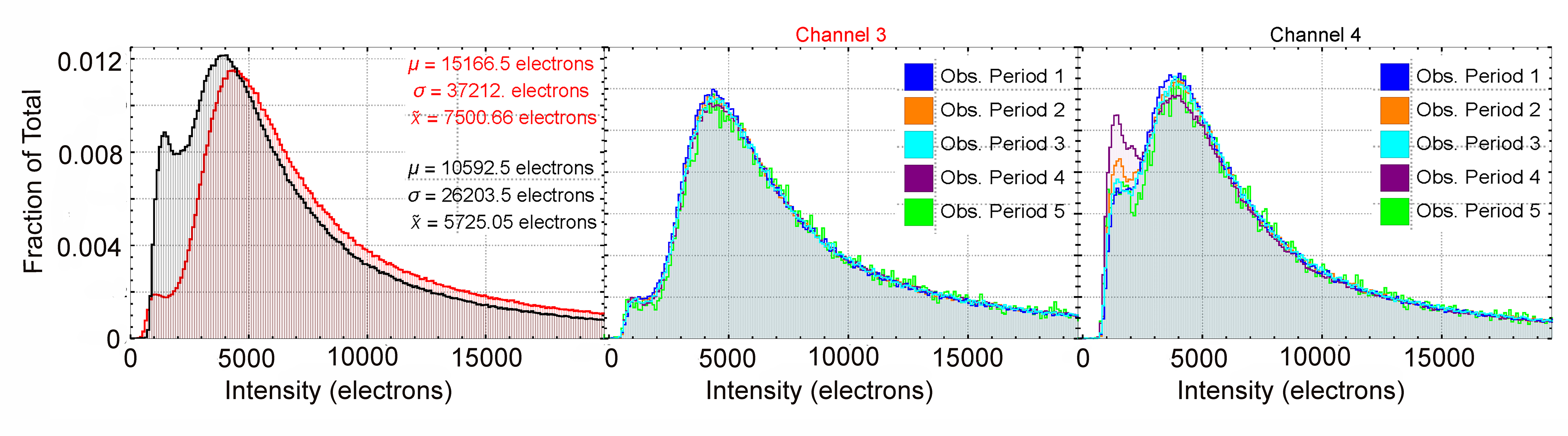}
\caption{Intensity profiles of the summed values of all pixels within isolated groups of radiation hits.}
\label{fig32}
\end{figure}

\begin{figure}
\centering
\includegraphics[width=.85\textwidth]{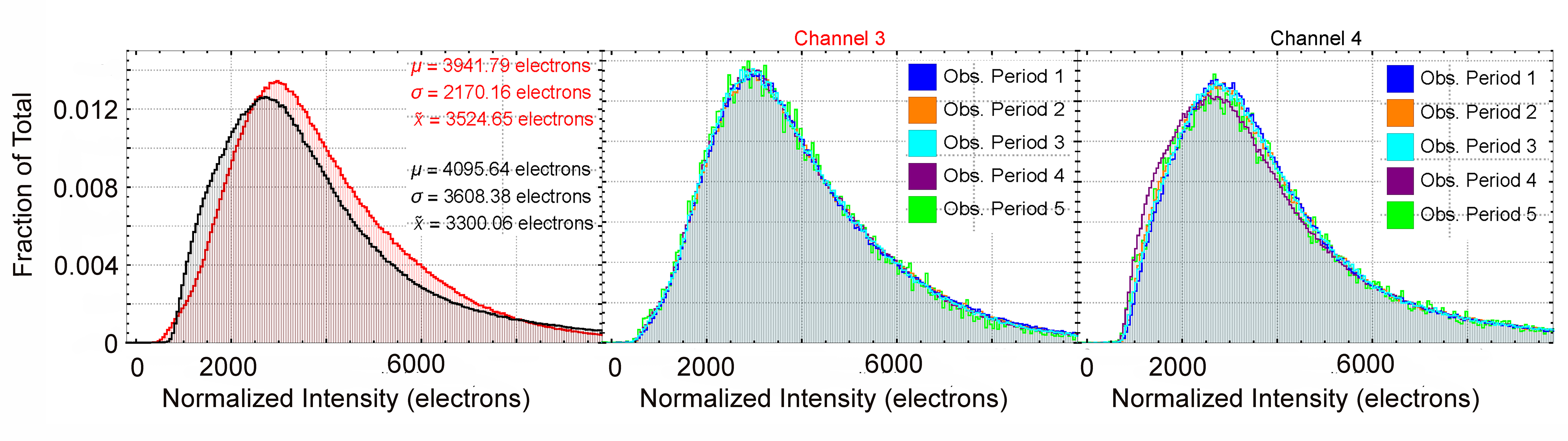}
\caption{Same as Figure \ref{fig32}, but the intensity is normalized by dividing by the number of pixels that comprise the isolated group. }
\label{normalized_mag_profile_radhits}
\end{figure}

\begin{figure}
\centering
\includegraphics[width=.85\textwidth]{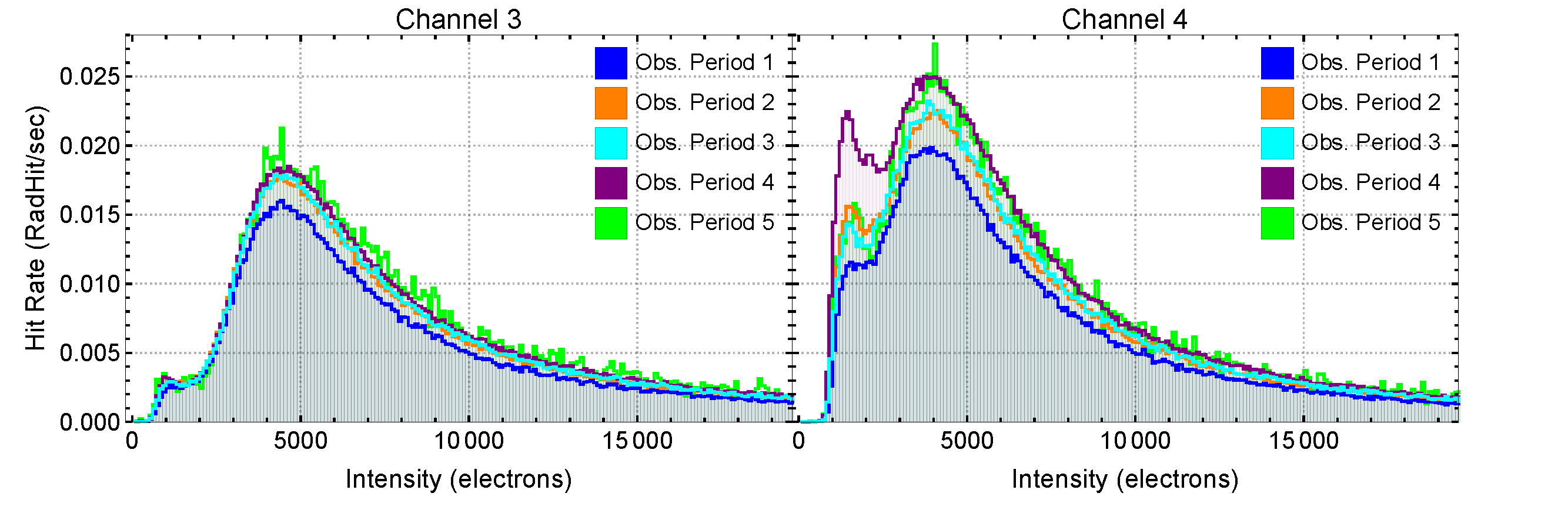}
\caption{Hit rate (RadHit/second) as a function of intensity.}
\label{rad_hit_rate_function_of_magnitude}
\end{figure}

\begin{figure}
\centering
\includegraphics[width=.85\textwidth]{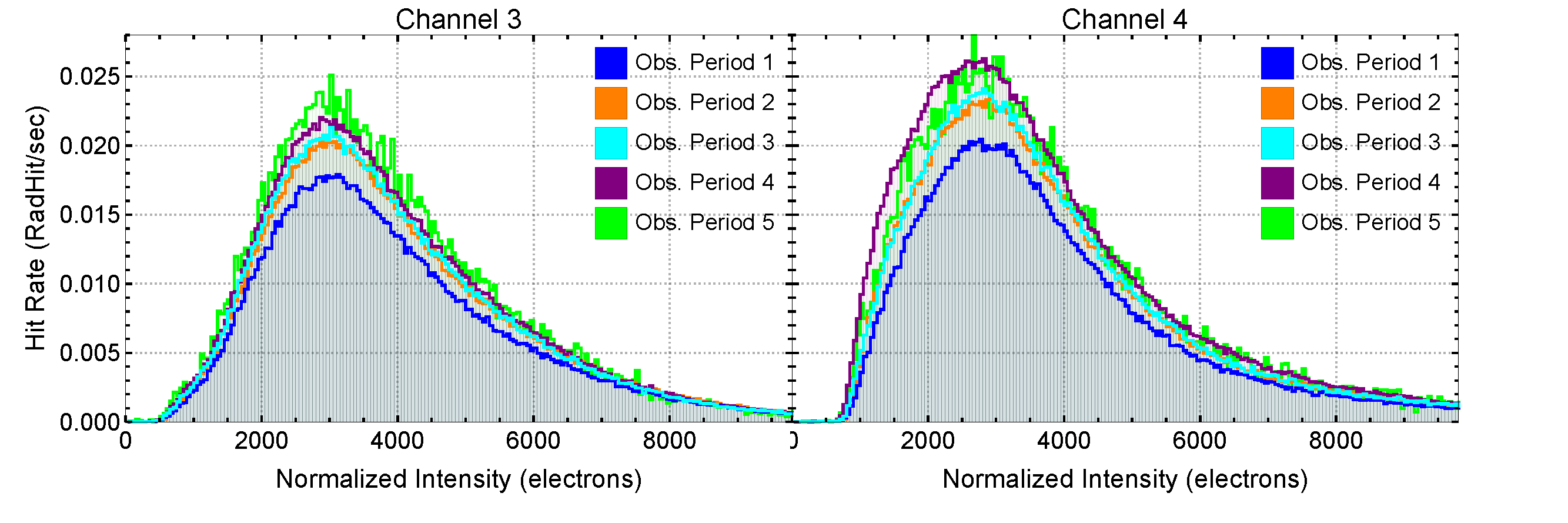}
\caption{Same Figure as \ref{rad_hit_rate_function_of_magnitude}, but the intensity is  normalized by dividing by the number of pixels that comprise the isolated group.}
\label{rad_hit_rate_function_of_magnitude_normalized}
\end{figure}

We next perform a similar analysis with the amplitude distributions of the isolated groups to probe the entire yields from a cosmic ray hit. Figure \ref{fig32} plots the intensity for each group summed over all the pixels. Figure \ref{normalized_mag_profile_radhits} normalizes these numbers dividing by the number of pixels in the group. In Figure \ref{fig32} the intensity profiles are similar for hits with higher intensity. The Channel 4 data contain more hits of lower intensity, and vary significantly in the different observation periods. This variation is due to hits that are one pixel in size, as previously noted in Figure \ref{montecarlo}. When we normalize the intensity of a group by the number of pixels in the group, the intensity profiles are similar overall.
Like Figure \ref{fig22}, Figures \ref{rad_hit_rate_function_of_magnitude} and \ref{rad_hit_rate_function_of_magnitude_normalized} plot the hit rate versus the hit amplitude, but for the groups of pixels. The shape of these profiles is similar to the intensity profiles found in Figures \ref{fig32} and \ref{normalized_mag_profile_radhits}.

\begin{figure}
\centering
\includegraphics[width=.85\textwidth]{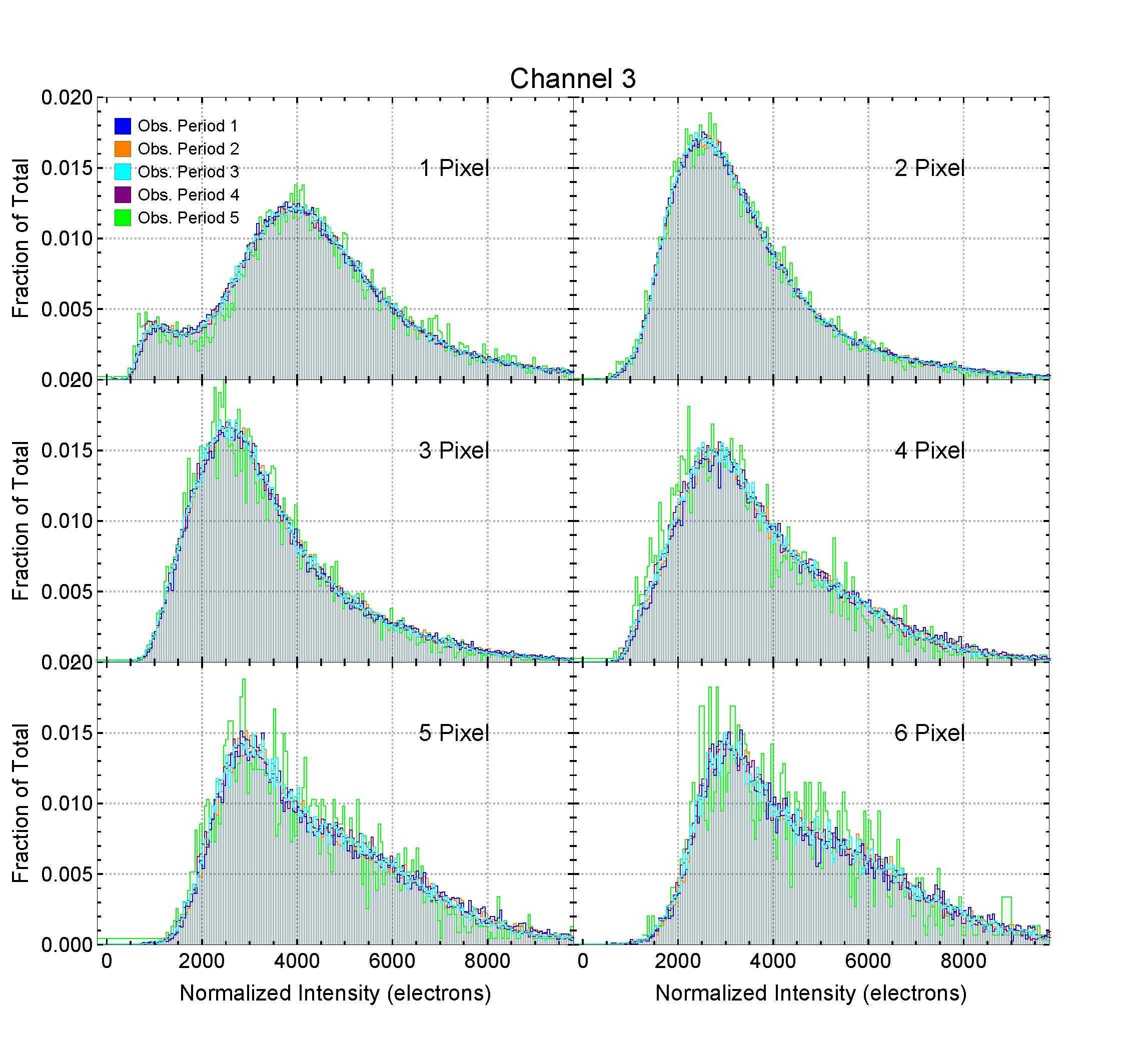}
\caption{Intensity distributions for hits of various pixel groupings for Channel 3. The intensity distributions remain consistent over time for each group size.}
\label{ch3_magnitude_profile_by_pix_size_and_time_period}
\end{figure}

\begin{figure}
\centering
\includegraphics[width=.85\textwidth]{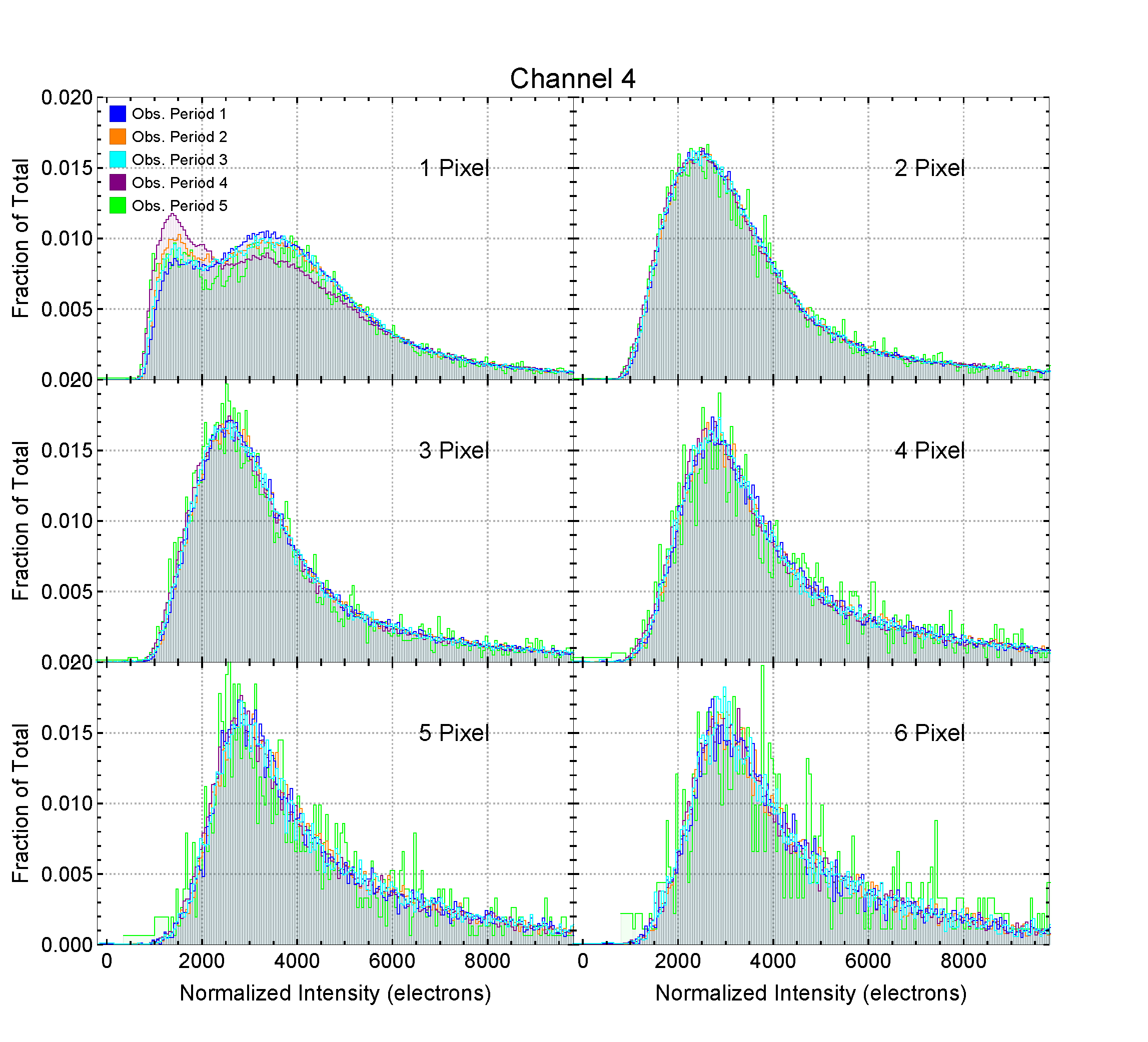}
\caption{Similar to Figure \ref{ch3_magnitude_profile_by_pix_size_and_time_period} but for Channel 4. The intensity distribution remain consistent over time for different group sizes, with the exception of the single pixel hits. Note that the first mode for the single pixel hit varies the most.}
\label{ch4_magnitude_profile_by_pix_size_and_time_period}
\end{figure}

Finally, we compare the intensity distributions for different group sizes across the five observation periods (Figures \ref{ch3_magnitude_profile_by_pix_size_and_time_period} and \ref{ch4_magnitude_profile_by_pix_size_and_time_period}). The intensity profiles do not change significantly between observation periods, with the exception of the single pixels in Channel 4.  This lack of variations reflects both the relatively weak diagnostics of the GCR spectrum through these intensity profiles, and that the GCR spectrum does not change dramatically with time in the relevant 30 - 300 MeV range \citep[e.g.,][]{buchvarova2010}. For single-pixel events in channel 4,  {``}Observation Period 4{''} (shown in purple) has the strongest first mode, while {``}Observation Period 1{''} (shown in blue) has the weakest bimodal behavior. This difference in the face of the lack of any other differences supports the suspicion of a weakly unstable spiking behavior contributing single-pixel counts in this channel. 

\subsection{Crosstalk} 
\label{crosstalk_sec}

\begin{figure}
\centering
 \includegraphics[width=.6\textwidth]{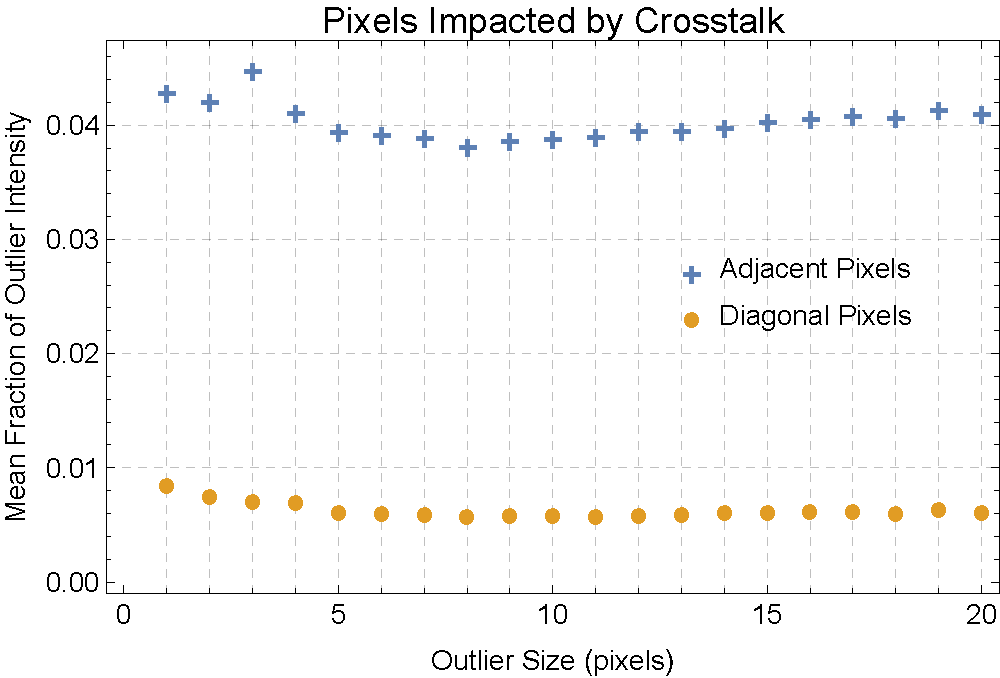}
 \caption{Intensity of the pixels neighboring flagged pixels with hits. We take the mean value of all adjacent pixels and diagonal pixels and divide by the average intensity of the hit (i.e., central) pixels.}
 \label{crosstalk}
\end{figure}

Figure \ref{crosstalk} plots the values of the adjacent and diagonal pixel hit intensity relative to that of the central pixel.  The average value for adjacent pixels is approximately 4.3\% of the average for the hit pixel, whereas the diagonal pixels received only 0.7\%.  These values are comparable for hits regardless of intensity. All together, this result is indicative of some form of crosstalk.  We note, however, that most of these adjacent pixels are not flagged as outliers because they are below the threshold for identification as separate hits. 

There are two plausible mechanisms for crosstalk: (1) distribution of free charge carriers from one pixel to another; and (2) interpixel capacitance, IPC; \citep[e.g.,][]{finger2005,fox2008,rauscher2007}. Models of these detectors indicate that the diffusion length is $\sim$ 2.5 $\mu$m \citep{rieke2015}, which applies to thermalized free electrons.  In addition, ionization by protons of $\ge$ 100 MeV can produce super-thermal free electrons that migrate much further, e.g., to 10 $\mu$m \citep[e.g.,][]{han2017}. Crosstalk by motions of free electrons will be a statistical process. In comparison, interpixel capacitance IPC), unavoidable between the members of the grid of detector contacts, should be a purely electrical phenomenon and should not depend on the type of particle producing the hit (e.g., proton or electron or muon), or its hit position within a pixel.

IPC is a cause of crosstalk for all infrared arrays, but we need to determine whether motion of free electrons also contributes. To do this, we identified cases of muon hits on a MIRI detector array isolated to a single pixel. To remove instances where the muon had passed through a small portion of a second pixel, we rejected the adjacent pixel with the largest effect from the hit. Using the three remaining adjacent pixels we compared the mean frame differences to the frame differences of the pixel with the muon hit. A low level of crosstalk (Dcross $\sim 3.06 \pm 0.66$\%) was found for pixels adjacent to one hit by a muon, consistent with the results for the IRAC detectors given the somewhat different detector architectures. However, there was substantial rms scatter (more than 20\% of the average value), inconsistent with the deterministic nature of IPC-driven crosstalk. We conclude that, on top of the crosstalk due to IPC, there is a significant contribution from thermal and/or super-thermal electrons migrating between pixels. 


\begin{figure}
\centering
 \includegraphics[width=.6\textwidth]{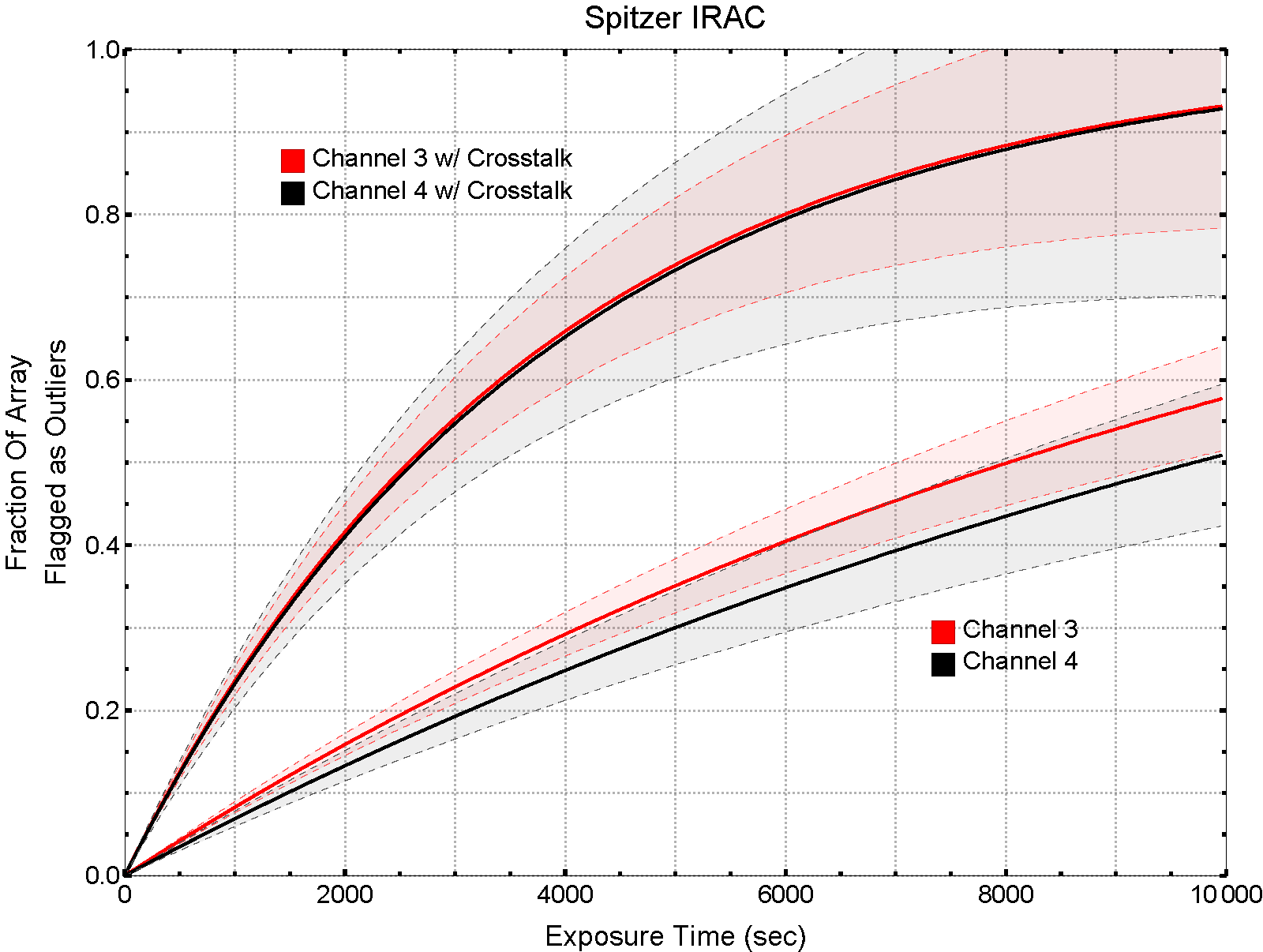}
 \caption{Fraction of the Spitzer IRAC array flagged as outliers over time, given the average hit rates for Channel 3 (in red) and Channel 4 (in black).}
 \label{max_exposure_fig}
\end{figure}

Figure \ref{max_exposure_fig} shows the hit rates, parameterized as the fraction of the total number of pixels that are flagged as outliers as a function of time.  After adding pixels affected by crosstalk rather than being hit directly, the number of pixels affected by a hit increases over the single-pixel estimates by more than a factor of three in both IRAC Channels. The hit rate itself would have a linear dependence as shown for the shorter time intervals, e.g., $\sim$ 7.5\% of the pixels hit in 1000 seconds. The nonlinearity for longer times results, however, because already-hit pixels are not flagged when they are hit again.


\section{Conclusions}
\label{sec:conclusion}

We have studied the effects of ionized particle hits on the Spitzer/IRAC Si:As IBC detectors. Although based on this single detector type, our results should have general applicability to solid state visual and infrared (and X-ray) detectors operated in space. We find behavior that matches expectations for hits at and above the minimum average ionization yield. Specifically, (1) from the minimum average yield level to a few times higher, the distributions of the numbers of ionized particles agree well between the two IRAC detector arrays and also with the ISO SWS Si:As IBC detectors;  (2) In contrast, the JWST MIRI detectors operated on the ground and hence, being hit primarily with single atomic charge muons, follow the Landau Distribution, i.e. the theoretical relation expected accounting for the small number of ionization events where substantial energy is transferred; (3) the number of hits yielding $\ge$ 10,000 electrons is much higher for the detectors in space than for the JWST MIRI Si:As IBC devices, reflecting the lack of multi-charged ionizing particles in the laboratory environment. Although the hit rates for the IRAC arrays track the ionizing proton flux in space, the overall hit rates in space are significantly higher (by factors of $\gtrsim$ 2) than predicted from models of the Galactic cosmic ray flux.  A similar excess has been reported for many other space-borne detectors, e.g. near infrared HgCdTe devices  \citep{ladbury2002} and CCDs used as X-ray detectors \citep{predehl2021}. In addition, nearly half of the events are below the minimum average ionizing yield ($\sim$3000 electrons).

\citet{garcia2018} also found a large number of low-intensity events when analyzing cosmic rays in the Gaia CCDs.  They could not reproduce this tail with their initial analytical model, which had an overly simplistic energy deposition description.  When using a more accurate Monte-Carlo simulation using GRAS \citep{santin2005}, a space environment effects analysis tool that takes into account the spacecraft geometry, a broader set of particle species, particle-matter interaction processes, and the statistics of the ionization process,  they were able to model the low intensity portion of the curve accurately.

Our conclusions are:

\begin{itemize}
    \item Without further information, it is prudent to anticipate ionizing particle hit rates higher than the predictions from CREME96 models by a factor of 2 $-$ 2.5$\times$.
    \item Although there are many hits with intensities below the intensity of a minimum ionizing energetic particle, the size distribution drops rapidly below $\sim$700 electrons.  In other words, there are few hits so small they would be difficult to detect with modern solid-state detector arrays.
    \item The number of pixels affected by a hit is dominated by: (1) the angle of incidence of the cosmic ray and hence how many pixels it passes through; and (2) interpixel capacitance that couples the free charge generated in the hit to neighboring pixels. There is a small amount of additional spreading on average due to: (1) scattering and showering of energetic particles generated by the hit; and (2) the increased migration distance of some of the ionization products that have been given significant energy by the hit.
    \item It should be possible to mitigate nearly all cosmic ray hits, given the read noises achieved with state-of-the-art detector arrays. Examples of the necessary approaches are provided by \citet{anderson2011,zhang2019}.
    \end{itemize}

\section{Acknowledgements}

This research has made use of the NASA/IPAC Infrared Science Archive, which is funded by the National Aeronautics and Space Administration and operated by the California Institute of Technology. This work is based on observations made with the Spitzer Space Telescope, which is operated by the Jet Propulsion Laboratory, California Institute of Technology under a contract with NASA. The authors would like to thank the Spitzer Help Desk personnel for assisting in this investigation - specifically Seppo Laine, Sean Carey, and Patrick Lowrance; they provided valuable guidance and insight to our process. A special thanks to Rene Gastaud, who also provided some valuable insight into this analysis. Ray Ladbury has also provided insightful comments. We also thank Cian Crowley for very helpful advice on interpreting the hit rates on the Gaia CCDs. Jane Morrison and Andras Gaspar contributed to the analysis of crosstalk from muon hits on MIRI arrays. We thank the anonymous referee for a very detailed and knowledgeable report.  


\appendix

\section{Determination of Proton Fluxes}
\label{appenda}

\begin{figure}
\centering
 \includegraphics[width=.7\textwidth]{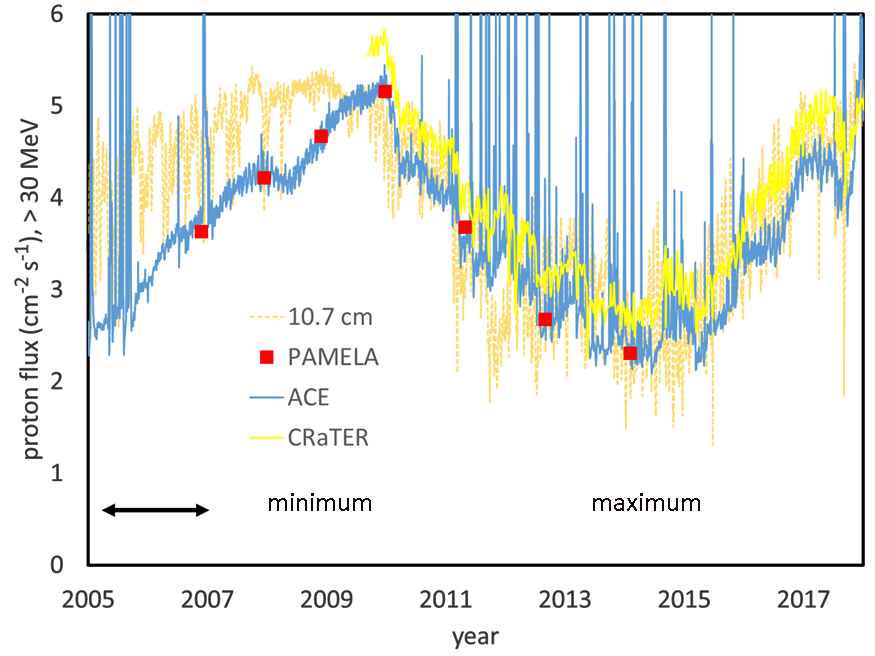}
 \caption{Determinations of the proton flux for energies $>$ 30 Mev. The double arrow at the lower left is the period during which the data were obtained that we analyze in this paper. The minima and maxima of solar activity are also indicated along the bottom of the figure.}
 \label{fig1a}
\end{figure}

Comparison of detector hit rates with the corresponding proton fluxes is often performed using nominal proton fluxes without regard to solar cycle. Figure~\ref{fig1a} shows the GCR flux in the relevant energy range over a full 11-year (and slightly extra) Solar Cycle (based on NOAA 10.7 cm measurements and proton data from \citet{adriani2013,martucci2018} (PAMELA), \citet{reitz2012} (CRaTER) and the Advanced Composition Explorer (ACE)). The flux varies by about a factor of two in anti-synchronism with the Solar Cycle (e.g., the minimum for Solar Cycle 24 began at the end of 2008 and the maximum in early 2014). Failure to account for these variations can result in inconsistencies in the hit rates relative to the proton fluxes. The double arrow at the lower left indicates the interval over which the data were obtained for this paper, which is near the maximum for Solar Cycle 23.

Accurate comparison with detector behavior requires a high cadence (i.e., weekly) measurement of the proton fluxes, to match their variations (see Figure~\ref{fig3}). The Solar radio flux at 10.7 cm is often used as a high cadence indicator of the level of activity and can be inverted to give a corresponding measure of the expected proton flux \citep[e.g.,][]{mendoza2014}, which we have done in Figure~\ref{fig1a}. We tested the applicability of these results by comparing with direct proton flux measurements obtained with the PAMELA instrument. Differential fluxes from PAMELA are provided for energies starting at 82 MeV  \citep{adriani2013,martucci2018}, which we have integrated. Nominal shielding for spaceflight instruments removes protons with energies $\lesssim$ 30 MeV \citep{glasse2020}. We therefore made a small correction (4.8\%) to give an equivalent threshold of $\sim$ 30 MeV.\footnote{The correction to $\sim$ 30 MeV was determined from the CREME96 spectrum in \citet{barth2000}.} The small size of this correction illustrates that the fluxes at a shielded detector should not be very sensitive to the details of the shielding\footnote{Determining ranges from the NIST web site \url{https://physics.nist.gov/PhysRefData/Star/Text/PSTAR.html}, the difference in predicted hit rates for shielding ranging from 5 to 20 mm of aluminum is only a few percent.}. As shown in Figure~\ref{fig1a}, there is a substantial difference between the estimates from the 10.7 cm flux and the direct proton flux measurements in the period 2007 - 2009, and the trend indicates that this discrepancy may persist throughout the period when we analyzed the Spitzer data, indicated by the double arrow.

The measurements with PAMELA do not have an adequate cadence for our purposes,  so we turned to measurements of the $>$ 30 MeV proton flux from the Advanced Composition Explorer (ACE). ACE was launched on August 25, 1997 and orbits the Sun-Earth L1 Lagrange point. The mission objective is to determine the elemental and isotopic composition of the cosmic rays, and compare that with similar data for the Solar wind.\footnote{\url{ http://www.srl.caltech.edu/ACE/ASC/index.html}}

ACE has a number of advantages for our purposes. It is not in Earth orbit, it provides data over two full Solar cycles, and its Solar Isotope Spectrometer and Cosmic Ray Isotope Spectrometer (SIS-CRIS) instrument T4 count rates are dominated primarily by protons of energies $>$ 10 Mev. We have used the data provided for an energy threshold of 30 MeV. However, these proton fluxes are available  only as ``browse quality'' or Level 1. Their calibration is uncertain, and there are concerns about contamination from scattered particles. Since the focus of the ACE project is on more massive elements, the proton data have not been made available with higher level processing. We therefore verified these data for our purposes as follows. We averaged the results over the two Solar cycles and normalized the average to the average CREME96 isotropic cosmic ray flux of 4.1 cm$^{-2}$ sec$^{-1}$ from \citet{evans2000}. As shown in Figure~\ref{fig1a}, the resulting calibrated ACE fluxes agree nearly perfectly with the measurements from PAMELA (although the calibration of the ACE measurements was done completely independently from PAMELA).

We used a different set of data to verify the overall behavior further. This second comparison is from CRaTER, the Cosmic Ray Telescope for Effects of Radiation, on the Lunar Reconnaissance Orbiter \citep{reitz2012}. The threshold for these measurements is 180 MeV; a 12\% correction was applied to derive the equivalent with a threshold of $\sim$ 30 MeV.\footnote{Again from the spectrum in \citet{barth2000}.}  As shown in Figure~\ref{fig1a}, these measurements track the calibrated ACE fluxes well, with a slight ($\sim$ 10\%) offset toward overall higher values. This offset might arise in part from the relatively large difference in energy thresholds (182 MeV vs. 30 MeV), i.e., it might be a small calibration difference associated with spectral variations over the Solar cycle \citep{buchvarova2010}. Nonetheless, the comparison confirms that the calibrated ACE measurements accurately track the full range of GCR flux variations. These results also show that between 2010 and 2018 the fluxes deduced from the 10.7 cm measurements are in close agreement with the direct measurements.

The calibrated ACE proton fluxes are plotted in Figure~\ref{fig3} and used to compare with the hit rates on the IRAC detectors in the main body of this paper.

\end{document}